# Thermal Tides in the Martian Atmosphere near Northern Summer Solstice Observed by ACS/TIRVIM onboard TGO

**Siteng Fan[1], Sandrine Guerlet[1,2], François Forget[1], Antoine Bierjon[1], Ehouarn Millour[1], Nikolay Ignatiev[3], Alexey Shakun[3], Alexey Grigoriev[4], Alexander Trokhimovskiy[3], Franck Montmessin[5], Oleg Korablev[3]**

[1]LMD/IPSL, Sorbonne Université, PSL Research Université, École Normale Supérieure, École Polytechnique, CNRS, Paris, France

[2]LESIA, Observatoire de Paris, CNRS, Sorbonne Université, Université Paris-Diderot, Meudon, France

[3]Space Research Institute (IKI), Moscow, Russia

[4]Research School of Astronomy and Astrophysics, Australian National University, Canberra, Australia

[5]LATMOS/IPSL, Guyancourt, France

Corresponding author: Siteng Fan (sfan@lmd.ipsl.fr)

**Key Points:**

- Thermal tides in the Martian atmosphere are investigated using temperature profiles retrieved from TIRVIM nadir observations.
- Diurnal tide dominates daily temperature variations; semi-diurnal tide and diurnal Kelvin wave are also important.
- Observations agree well with numerical simulations, but suggest phases of diurnal and semi-diurnal tides earlier than predicted.

**Abstract**

Thermal tides in the Martian atmosphere are analyzed using temperature profiles retrieved from nadir observations obtained by the TIRVIM Fourier-spectrometer, part of the Atmospheric Chemistry Suite (ACS) onboard the ExoMars Trace Gas Orbiter (TGO). The data is selected near the northern summer solstice at solar longitude ($L_S$) 75°-105° of Martian Year (MY) 35. The observations have a full local time coverage, which enables analyses of daily temperature anomalies. The observed zonal mean temperature is lower by 4-6K at ~100Pa, but higher towards the summer pole, compared to the LMD Mars General Circulation Model (GCM). Wave mode decomposition shows dominant diurnal tide and important semi-diurnal tide and diurnal Kelvin wave, with maximal amplitudes of 5K, 3K, and 2.5K, respectively, from tens to hundreds of Pa. The results generally agree well with the LMD Mars GCM, but with noticeable earlier phases of diurnal (~1h) and semi-diurnal (~3h) tides.

**Plain Language Summary**

Unlike the Earth, daily temperature variation on Mars is as large as tens of degrees because of its thin atmosphere. The sunlight absorbed by the Martian surface and dust in the atmosphere leads to dramatic temperature increase in the lower atmosphere during daytime. Such large and regular changes can trigger temperature waves, and some modes of them can propagate into higher altitudes, where they become the major factor controlling the daily temperature variation. In this work, temperature profiles obtained using thermal-infrared spectra are analyzed. Zonal mean temperature is compared with numerical simulations of the Martian atmosphere. Different types of the wave mode are computed through decomposition. Results from the observation agree well with model prediction when the observation mechanisms are taken into consideration. Estimation of the strength of these waves can be improved in the future with improved design of observation strategies.

**1 Introduction**

Atmospheric thermal tides are planetary-scale harmonic responses driven by diurnal solar forcing and influenced by planetary topography (Gierasch & Goody 1968, Zurek 1976). As results of solar heating absorbed directly by the atmosphere and exchanged with the surface, thermal tides in the Martian atmosphere usually have large amplitudes due to its low heat capacity, and dominate the diurnal temperature variations throughout the thin atmosphere. Among the excited tides, some modes can vertically propagate into the middle and upper atmospheres with amplitudes growing exponentially with decreasing air density, which therefore influences the mean atmospheric circulation.

Diurnal temperature variations and related pressure changes have been observed by many Mars landers and orbiters (e.g., Hess et al., 1977, Banfield et al. 2000, Lee et al. 2009, Kleinböhl et al. 2013, Forbes et al. 2020), which have significantly enriched our knowledge of the Martian thermal tides in the past two decades. However, many recent Mars missions have had sun-synchronous orbits, where temperature observations are mostly limited to two local times which differ by half of a Martian solar day (sol), e.g., 2h and 14h for the Thermal Emission Spectrometer (TES) onboard Mars Global Surveyor (MGS, Banfield et al. 2000), and 3h and 15h for the Mars Climate Sounder (MCS) onboard the Mars Reconnaissance Orbiter (MRO), despite some offsets of ±1.5-3.0 hours of cross-track observations (Kleinböhl et al. 2013, Forbes et al. 2020). Mars Express has been in a non-solar-synchronous polar orbit, but its large orbital inclination only allowed it to monitor the diurnal cycle using the Planetary Fourier Spectrometer (PFS) within a long period comparable to a Martian year to cover all local times (Giuranna et al. 2021). Analyses using sun-synchronous observations can reasonably investigate the diurnal tide, while constraints on aliasing waves, especially those with even number in temporal frequency, require plenty of a priori knowledge about the latitude-pressure structure of each wave mode predicted by the tidal theory (e.g., Guzewich et al. 2012). Therefore, data with large local time and spatial coverage are required for analysis of such planetary-scale diurnal/sub-diurnal variations. Observations by TIRVIM can meet these requirements, which is the subject of this work.

## 2 Observations and Data Processing

### 2.1 Instrument and Observations

TIRVIM is a double-pendulum Fourier-spectrometer, one of the three instruments of the Atmospheric Chemistry Suite (ACS) onboard the ESA-Roscosmos mission ExoMars Trace Gas Orbiter (TGO, Korablev et al. 2018, Svedhem et al. 2020). The spacecraft is in a near-circular orbit with a radius of ~400km and a period of ~2 hours (Capderou & Forget 2004). The orbit is not sun-synchronous, and the sub-spacecraft point slowly drifts in local time (Figure 1a). A full diurnal cycle can be sampled in nadir viewing geometry between 74°N and 74°S within ~55 sols, which is equivalent to ~30° in solar longitude ($L_S$). Details of the TIRVIM instrument are presented in Korablev et al. (2018). Through nadir sounding, TIRVIM is designed to monitor the temperature in the Martian atmosphere and of the surface, as well as column integrated aerosols (dust and clouds). Under this observation scheme, the instrument is sensitive to thermal infrared from 620 to 1300cm$^{-1}$ with a spectral resolution of 1.2cm$^{-1}$. TIRVIM operated from April 2018 to December 2019, nearly one Martian year (MY) from MY34 $L_S$=134° to MY35 $L_S$=115°. It provided ~1.7 million thermal infrared spectra with decent data quality. Guerlet et al. (2022) presented a retrieval process based on optimal estimation theory that simultaneously retrieves atmospheric temperature profile, surface temperature, and the optical depths of dust and water ice clouds from these spectra. It builds initial guess of the temperatures using different parts of TIRVIM spectra (660-740cm$^{-1}$ for atmospheric and 1240-1290cm$^{-1}$ for surface temperature), and then it iterates with a radiative transfer model until the synthetic spectrum converges to the observed one or the maximal number (10) of iterations is reached. The derived temperature profiles typically cover a vertical range from ~2-3km above Mars' surface to ~50-55km (~1-2Pa) with a vertical resolution of ~1 scale height (10km) and a typical retrieval error of ~2K at pressure >30Pa. These temperature profiles cover most local time within sub-seasonal scales (Figure 1b), which enables the investigation of thermal tides on a sub-daily basis and resolves the harmonic aliasing.

### 2.2 Data Processing

Temperature profiles used for analysis in this work are selected near the northern summer solstice during $L_S$=75°-105° of MY35, which totals ~9.5×10$^4$ in number (Guerlet 2021). This is a clear season with no dust storms identified, so the temperature structure of the atmosphere is not expected to undergo drastic changes. As the observations are not evenly distributed at different local times and locations (Figure 1b), individual profiles are first binned with grid sizes of 5°, 10°, and 1h in longitude, latitude, and local time, respectively, to derive zonal-mean daily temperatures, and corresponding daily anomalies which are shown in Section 3.2. Each bin is assumed to have the same weight so that the sampling bias in local time and location can be decreased. In the wave mode decomposition (Section 2.3 and 3.3), however, only the latitude bin is included as the local time and the longitude are considered directly.

### 2.3 Wave Mode Decomposition

To investigate the contributions of each wave mode, temperature at a given latitude and pressure level is assumed to be a linear combination of waves as shown in Equation (1), and then different modes of waves could be decomposed accordingly using least-square fit, in line with previous works (e.g., Banfield et al. 2000, Lee et al. 2009., Wu et al. 2015 and 2017).

$$T(\lambda, \varphi, p, t) = \sum_{\sigma,s}\left(C_{\sigma,s}(\varphi, p)\cos(s\lambda + \sigma t) + S_{\sigma,s}(\varphi, p)\sin(s\lambda + \sigma t)\right) \quad (1)$$

where λ, φ, and p are longitude, latitude, and pressure level, respectively; t is the universal time; s and σ are the frequencies in longitude and time; C and S are the coefficients of the cosine and sine functions at a certain latitude and pressure level, which are the unknowns in the decomposition. Positive values of s/σ denote westwards propagating waves; e.g., (s, σ) = (1, 1) indicates the mode with wavenumber one in longitude and period of one Martian day in time, which is the diurnal tide. Temperature profiles are binned in latitude, and within each latitude bin individual temperature profiles are considered with the same weight in the decomposition to compute the corresponding coefficients (C and S) of this latitude bin. Applying the decomposition to the three-dimensionally binned data has similar results if each bin is weighted by corresponding observation numbers, but using individual profiles leads to more precise values of time and location. The truncation of the expansion is selected as s={0, 1, 2, 3} and σ={-2, -1, 0, 1, 2}, which is the same as Wu et al. (2015 and 2017). This results in 35 linearly independent unknowns of C and S at each pressure level in a given latitude bin, which typically consists of thousands of temperature observations. Higher order waves are ignored as their contributions are thought to be negligible. Amplitude (A) and phase (θ) of each mode can finally be obtained by combining the coefficients.

$$A_{\sigma,s}(\varphi, p) = \sqrt{C_{\sigma,s}(\varphi, p)^2 + S_{\sigma,s}(\varphi, p)^2} \quad (2)$$

$$\theta_{\sigma,s}(\varphi, p) = \tan^{-1}\left(\frac{C_{\sigma,s}(\varphi,p)}{S_{\sigma,s}(\varphi,p)}\right) \quad (3)$$

## 3 Results

### 3.1 Temperature Profiles

Example temperature profiles obtained near 9h and 21h between ±5° in latitude are shown in Figure 1c. Profiles at these two local times show consistent differences. The atmosphere is warmer at 21h near the surface, while colder at ~100Pa. As a result of nadir sounding, temperature retrieved at a specific pressure level is a weighted average of a large vertical region, e.g., ~10-80Pa for temperature retrieved at 30Pa (Figure 1d), so the temperature profiles are smoother than those retrieved using limb sounding, e.g., MCS (Lee et al. 2009). However, differences among the temperature profiles obtained at these two local times are consistently monitored, as the vertical convolution does not vary much with local time (Figure 1d). More details and examples are discussed in Guerlet et al. (2022).

3.2 Zonal mean daily temperature

To investigate the zonal mean daily temperature, the binned data are averaged along the longitude dimension, then diurnal mean and anomaly are derived. Results of the observed diurnal mean temperature show a typical solstice distribution (Figure 2a), including an indication of warming at tens of Pa towards the winter pole due to the downwelling branch of the Hadley circulation. Prediction of such observation using the Laboratoire de Météorologie Dynamique (LMD) Mars Global Circulation Model (GCM, Forget et al. 1999, Fan 2022) is given in Figure 2b, which is computed by binning the outputs over $L_S=75°-105°$. Description of the GCM is given in the Supporting Information. The diurnal mean temperature distribution does not change much when the model outputs are sampled at the same locations and times as observations (Figure 2c), and with vertical convolution considered (Figure 2d). These two factors are important in the investigation of diurnal anomalies (see below). The GCM prediction shows a good agreement with observations, but with noticeable differences (Figure 2e), and these differences are more consistent when the observation scheme is considered (Figure 2f). Compared to the TIRVIM results, the model is warmer by 4-6K near 100Pa, and slightly colder above, while the temperature towards the summer pole is consistently underestimated with a difference of 4-6K from the surface to tens of Pa. These differences are likely due to the challenges of simulating physical processes in the GCM, e.g., dust and/or water clouds (Navarro et al. 2017), and are indications and constraints for further model improvement.

Daily temperature anomalies are obtained by subtracting the diurnal mean from temperature profiles. An example in the equatorial region between ±5° latitude is given in Figure 3a. It is the first time that such diurnal variations in the Martian atmosphere are observed at almost all local times within a Martian season. It shows a signature of diurnal tide with a downwards propagating anomaly of 2-4K, from approximately 20Pa at 0h to the surface at 16h (orange dashed line in Figure 3a). Another temperature maximum appears across all pressure levels near 15h, which is a sign of the semi-diurnal tide, but it is also likely introduced by the sampling scheme of TIRVIM. As the sub-spacecraft point slowly drifts in local time, and crosses each latitude twice per orbit (Figure 1a), sampling could superimpose a semi-diurnal cycle onto the temperature anomaly, as long as the background temperature keeps increasing/decreasing at certain latitudes within this season. In the case of TIRVIM observations, sampling at ~4h and ~16h after $L_S=100°$ (Figure 1a) can solely result in such a daily temperature anomaly even if there are no thermal tides.

Outputs of the LMD Mars GCM binned over $L_S=75°-105°$ are used for comparison (Figure 3b). The model prediction shows a similar downward anomaly phase propagation, and an above-average temperature at 30Pa in the early afternoon. The observation scheme is important in this investigation. When the model results are sampled at the same locations and times as the observations, the early morning temperature maximum becomes much larger (>10K), and the early afternoon anomaly extends to a much larger pressure range (Figure 3c). This agrees with the scenario that observation sampling amplifies the "apparent" semi-diurnal variation. Nadir sounding vertical convolution (Figure 1d), another factor introduced by observations, smooths temperature profiles and therefore the anomalies (Figure 3d). The final result shows good agreement with observations (Figure 3a) including the 2-4K "apparent" amplitude of the downwards propagating anomaly, and the anomaly in the early afternoon extending to all pressure levels. Nevertheless, two differences are noticeable. First, the model has a phase difference of the diurnal thermal tide (blue dashed line in Figure 3d), which is ~1h later than the observation. Second, the early afternoon anomaly extends more in the observation than the model, which is an indication of a larger amplitude and/or a phase difference of semi-diurnal tide than expected. This is assessed more quantitatively in Section 3.3.

### 3.3 Amplitude and Phase of Wave Mode

To investigate the contribution of wave modes, least-square fit using Equation (1) is applied to temperature observations at each latitude bin and pressure level, then amplitudes and phases of these wave modes are derived using Equations (2) and (3). An example of the decomposition in the equatorial region at 100Pa is given in the Supporting Information and Figure S1. Three wave modes are identified important: (1) the diurnal tide, the (1, 1) mode, which propagates westwards and sun-synchronously with wavenumber one in longitude, (2) the semi-diurnal tide, the (2, 2) mode, which also propagates westwards and sun-synchronously but with a wavenumber two in longitude, and (3) the diurnal Kelvin wave, the (1, -1) mode, which propagates eastwards with wavenumber one in longitude and with a period of one Martian day. Distributions of wave mode amplitudes and phases are then computed by applying the decomposition to each latitude bin and pressure level. The results are shown in Figure 4 for the retrieved TIRVIM temperature profiles (Figure 4a-4d), the original model outputs (Figure 4e-4h), the model outputs with the same sampling as observations (Figure 4i-4l), and with the same sampling as well as the vertical convolution (Figure 4m-4p). Consistency of the resulting amplitudes and phases indicates the success of the decomposition, as the implementation of decomposition is independent at each latitude and pressure level. Also, the derived background temperatures, the (0, 0) mode (the first column of Figure S2), have good agreements with their corresponding zonal mean values computed using binned data (Figure 2) without any visible differences.

The amplitude distribution of diurnal tide shows a good agreement between observations (Figure 4a) and the model (Figure 4e). At ~50Pa, it has a large maximal value (~5K) at ~50°N, and a smaller local maximum near the equator, while possible amplitude maxima at ~5Pa are smoothed out by the vertical convolution (Figure 4i and 4m). The overall structure of the observed phase distribution has a fairly good agreement with model prediction, both of which show consistent downward progression (Figure 4b and 4f). Noticeable differences are in the equatorial region at pressure levels less than ~30Pa. This is likely due to the small derived amplitude (Figure 4a), which results in the difficulty of determining the phase.

In contrast to the diurnal tide, the latitude-pressure structure of the semi-diurnal tide shows significant disagreements between observations and the model, except for the fact that they have similar maximal amplitudes of ~3K (Figure 4c and 4g). The observation scheme is important in this case. The shape of the structure becomes much similar when the model is sampled at the same locations and times as observations, but it meanwhile introduces larger "apparent" amplitude (Figure 4k). Vertical convolution smooths temperature gradients and therefore the derived tide amplitudes, which results in a much better agreement between the final model prediction (Figure 4o) and the observation (Figure 4c). This is consistent with the scenario described in Section 3.2. The phase of the semi-diurnal tide shows generally good agreements (the second column in Figure S2). Both observation and the model show downward phase progression, which is almost linear with the logarithm of pressure from 5 to 100Pa (Figure S3). Vertical wavelength of the semi-diurnal tide can therefore be derived by a linear fit to the phase progression, which is ~60km in the model if assuming a scale height of 10km, but it is ~90km when the observation scheme is considered. The observations show a good agreement in the tide wavelength, but it is earlier by

~π/4 (3h in local time), which potentially explains the different early afternoon anomaly shown in Figure 3.

Despite a smaller amplitude, the (1, -1) diurnal Kelvin wave is successfully distinguished. It has an amplitude maximum of ~2.5K at ~200Pa, indicated by both observations and the model (Figure 4d and 4h), and also a possible maximum at ~5Pa (Figure 4h), which is likely smoothed out by the vertical convolution (Figure 4p). The phase of the Kelvin wave also agrees between observations and the model (Figure S2c and S2o) if the observation scheme is taken into consideration. Interpretation of this mode is not influenced by the sampling scheme (Figure 4h and 4l), as it propagates eastwards, the other direction than sun-synchronous. Another mode, the (1, 0) stationary wave, is expected to be important at latitudes south of 45°S (Figure S2h), but it is not seen from the observations due to limited data coverage (Figure 1b).

**4 Discussion and Conclusion**

In this work, temperature in the Martian atmosphere is analyzed across multiple local times using profiles retrieved from nadir-viewing spectra obtained by TIRVIM during a dust-free season, $L_S$=75°-105° of MY 35. TIRVIM observations show dominant diurnal tide and important semi-diurnal tide and Kelvin wave, with influence from observation strategy. They agree with predictions made by the LMD Mars GCM, but with phases of diurnal and semi-diurnal tides earlier than expected.

Diurnal temperature variations in the Martian atmosphere are analyzed using observations with full local time coverage for the first time. No assumptions about the latitudinal distributions of atmospheric wave modes, or the Hough modes (Lindzen & Chapman 1969), whose characteristics depend on more assumptions about the physical properties of the Martian atmosphere, are required in the wave mode analysis. This was inevitable in previous works using observations from TES or MCS (e.g., Wilson 2000, Guzewich et al. 2012). Due to the full coverage of local time, phases of the wave modes are well constrained. The dominant diurnal tide has a phase varying from 0 to –(1/2)π within a wide pressure range (~50-200Pa, Figure 4b), which corresponds to its wave crest being in late morning. This suggests that the temperature differences computed at the sampling local times of TES and MCS (Banfield et al. 2003, Lee et al. 2009) within this pressure range are likely smaller than the actual tide amplitude. The amplitude of the semi-diurnal tide has a maximum near the equator and is observed to reach 2-3K in the lower atmosphere >5Pa, which agrees well with previous works during the same season (e.g., Banfield et al. 2000, Kleinböhl et al. 2013). However, the phase is earlier by ~3h than expected. Reasons leading to such a phase difference, as well as that for the diurnal tide, are currently unknown. Preliminary investigation using the GCM suggests that phases of modeled tides are not sensitive to parameters included in the current model, which indicates that the phase discrepancy is significant and presents a challenge to numerical simulations. The local time coverage also allows an investigation of the ter-diurnal tide, the (3, 3) mode, which is tested by expanding the truncation of σ to ±3 in Equation (1). This results in 49 linearly independent unknowns in the wave mode decomposition. However, the results suggest that this mode is not important, at least during this season, with an amplitude of <0.5K in most of the cases (Figure S4).

Two sources of degeneracy, vertical convolution and sampling scheme, largely influence the interpretation of thermal tides using TIRVIM observations. The vertical convolution originates from the nadir viewing geometry thermal sounding. It smooths the temperature vertical oscillations and therefore reduces the apparent daily anomaly. A comparison of these temperature profiles with those obtained from limb sounding by MCS (Kleinböhl et al. 2009) is presented in Guerlet et al. (2022), where two sets of profiles show good agreement when vertical convolution is included. This is consistent with the model-observation agreement shown in this work. The sampling strategy of local time is also important due to the slow drift of the observation local time (Figure 1a), which results in the modulation of seasonal temperature variations into daily temperature anomalies. As the seasonal changes can have large values, e.g., a ~2K increase in the equatorial region at 30Pa suggested by the GCM, a small difference in $L_S$ (~20° in the case of this work) could result in noticeable amplification of decomposed tides. This effect has larger influences on the interpretation of migrating than non-migrating tides, so the model-observation agreement is better for the diurnal Kelvin wave than the diurnal and semi-diurnal tides (Figure 4). Obtaining a full local time coverage within a short seasonal range is helpful to address this issue. Observations from the Emirates Mars Infrared Spectrometer (EMIRS, Edwards et al. 2021) onboard the Hope spacecraft are expected to provide promising results, with the capability of sampling all local times at a given region over 10 sols.

## Figures

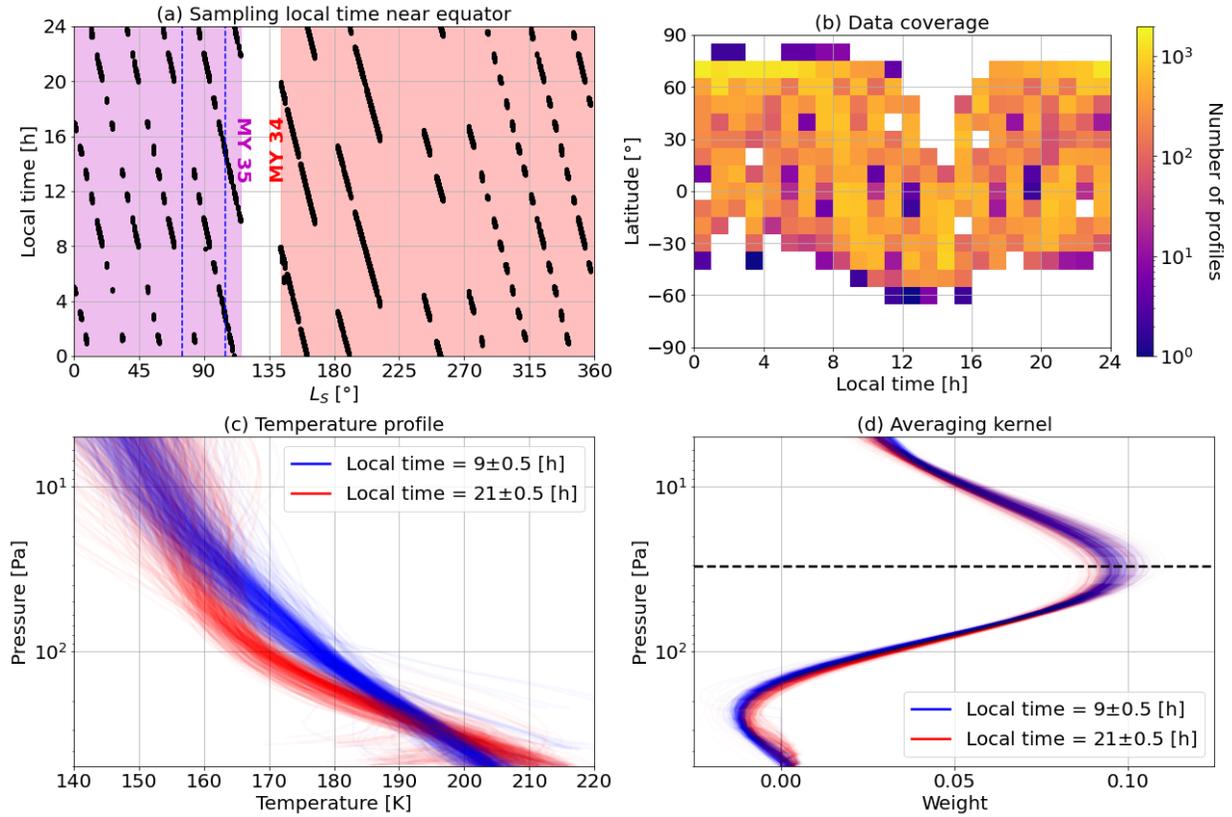

**Figure 1.** (a) Season and local time of TIRVIM temperature profile observations between ±5° in latitude (back dots). The two shaded areas denote MY 34 (red) and MY 35 (magenta). The two blue vertical dashed lines denote the season ($L_S$=75°-105°) selected in this work. (b) Numbers of TIRVIM temperature profiles in the (latitude, local time) bins. (c) Temperature profiles between ±5° in latitude and within half an hour of 9h (blue) and 21h (red) in local time. (d) Averaging kernels of retrieving the temperatures at 30Pa (black dashed line) using observations between ±5° in latitude and within half an hour of 9h (blue) and 21h (red) in local time.

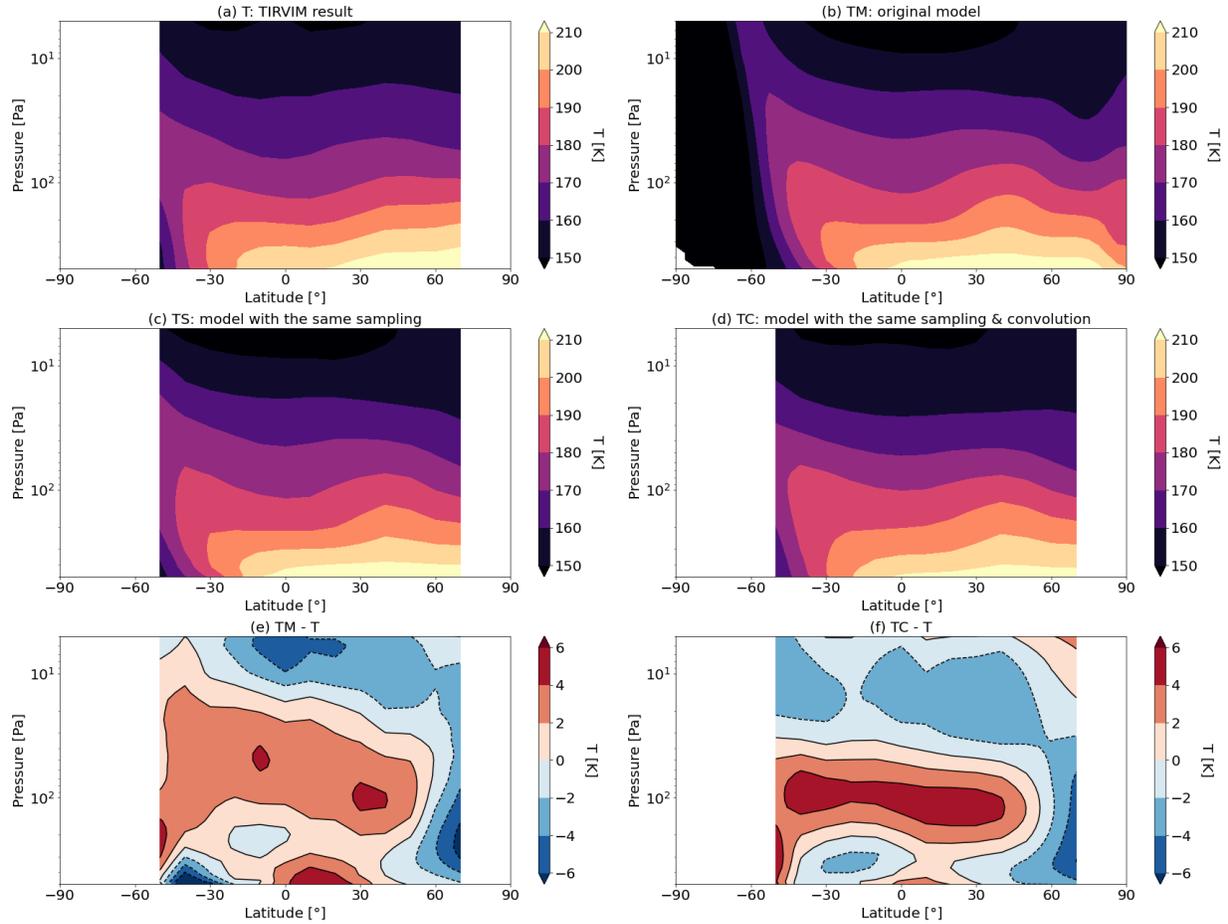

**Figure 2.** (a) Zonal and diurnal mean temperature computed using TIRVIM observations between ±5° in latitude during MY 35 $L_S$=75°-105°. (b) Same as (a), but for the LMD Mars GCM outputs. (c) Same as (b), but the outputs are sampled at the same locations and times as TIRVIM observations. (d) Same as (c), but it includes the vertical convolution. (e) Difference of the zonal and diurnal mean temperature between observations and the original GCM outputs, which is the difference between (a) and (b). (f) Same as (e), but for the difference between (a) and (d), where the GCM outputs are sampled and convolved the same way as observations.

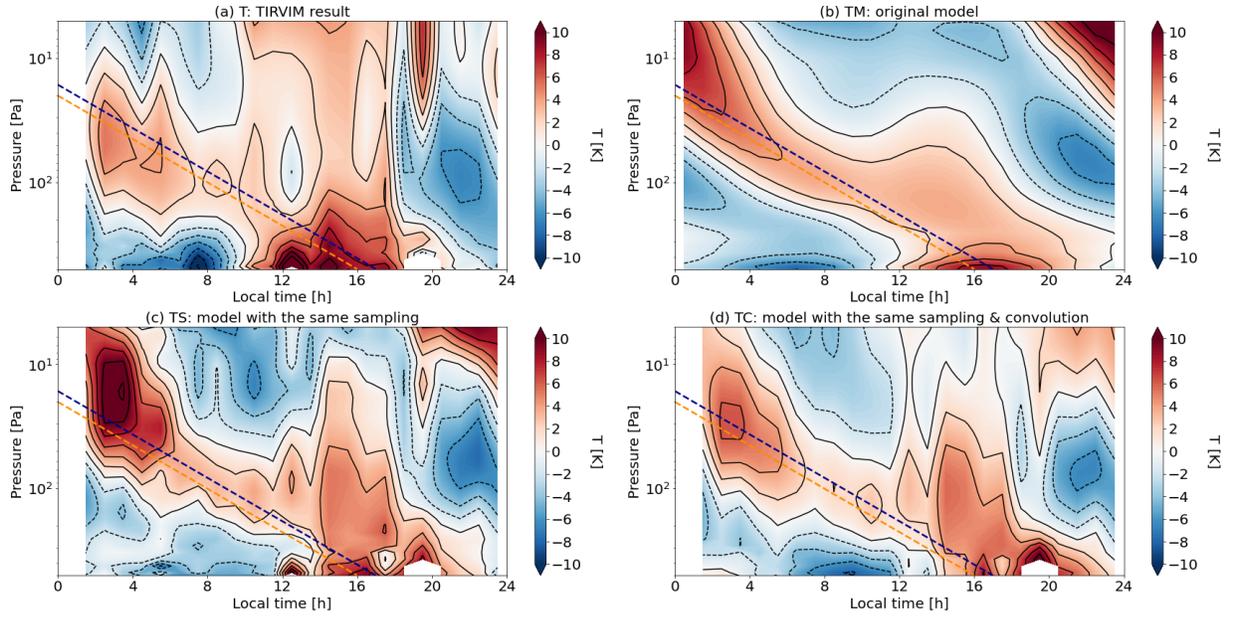

**Figure 3.** Same as Figure 2a-2d, but for zonal mean daily temperature anomaly. The yellow and blue dashed lines denote the approximate slope of the downward phase progression of temperature maximum for observations and model outputs, respectively, which differ by 1h in local time. Model-observation comparison should be done by comparing (a) and (d).

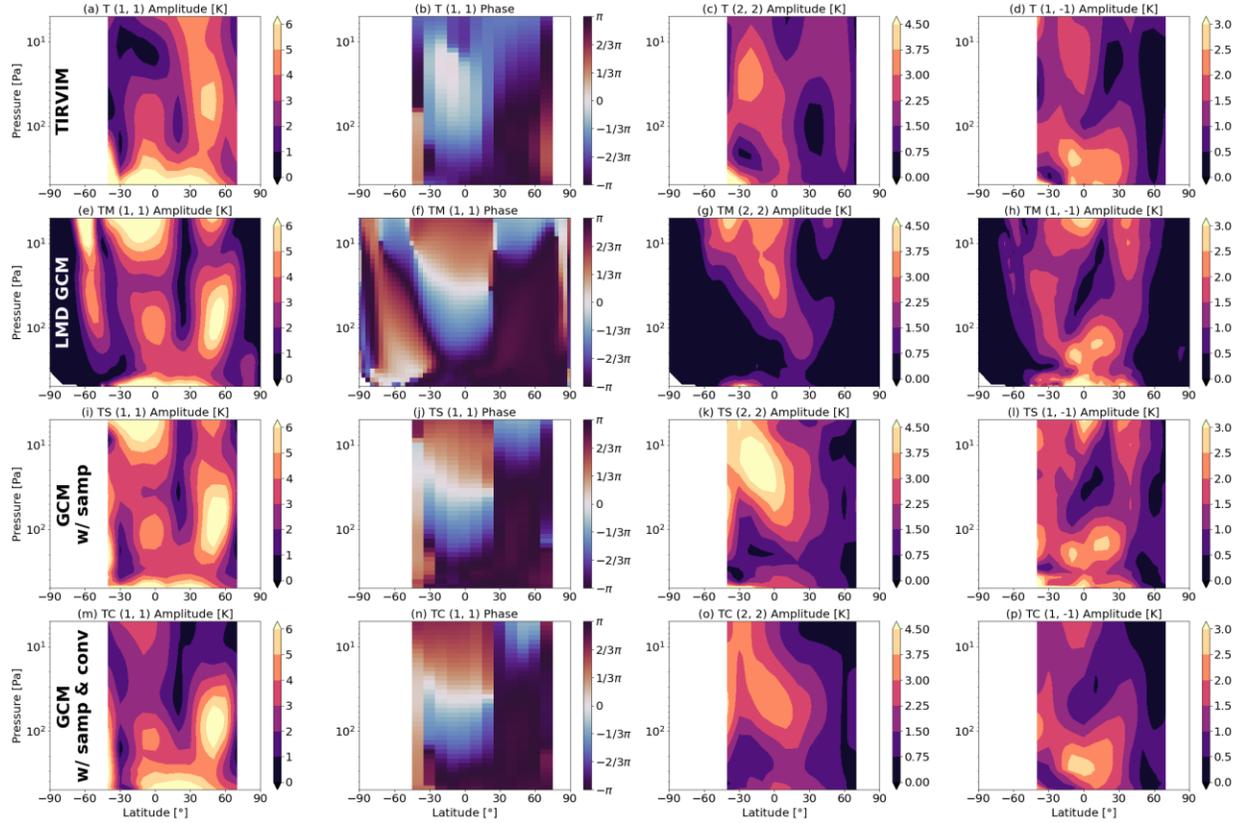

**Figure 4.** (a) Amplitude of the diurnal tide component derived from the wave mode decomposition using TIRVIM observations. (b) Same as (a), but for the phase of the diurnal tide. (c) Same as (a), but for the semi-diurnal tide. (d) Same as (a), but for the diurnal Kelvin wave. (e)-(h) Same as (a)-(d), respectively, but for the LMD Mars GCM outputs. (i)-(l) Same as (e)-(h), respectively, but for the model outputs sampled at the same locations and times as TIRVIM observations. (m)-(p) Same as (i)-(l), respectively, but it includes the effect of vertical convolution.

manuscript submitted to *Geophysical Research Letters*

**Acknowledgments**

This work was supported by CNES. It is based on observations with ACS/TIRVIM embarked on TGO. The science operations of ACS are funded by Roscosmos and ESA. The TIRVIM team at IKI acknowledges the subsidy of the Ministry of Science and High Education of Russia.

**Open Research**

The TIRVIM retrieval results used in this paper are available on the Institut Pierre Simon Laplace (IPSL) data server with doi: 10.14768/ab765eba-0c1d-47b6-97d6-6390c63f0197. The LMD Mars GCM output during $L_S$=60°-120° of MY 35 is also available on the IPSL data server with doi: 10.14768/9853d6ae-7d11-450f-9363-ac8cf7bd143a. Permission is granted to use these datasets in research and publications with appropriate acknowledgements that are presented on the dataset websites.

manuscript submitted to *Geophysical Research Letters*Wu, Z., Li, T., Dou, X. (2015). Seasonal variation of Martian middle atmosphere tides observed by the Mars Climate Sounder. *Journal of Geophysical Research (Planets),* 120(12), 2206-2223. https://doi.org/10.1002/2015JE004922

Zurek, R. W. (1976). Diurnal tide in the Martian atmosphere. *Journal of Atmospheric Sciences,* 33, 321-337. https://doi.org/10.1175/1520-0469(1976)033<0321:DTITMA>2.0.CO;2

# Thermal Tides in the Martian Atmosphere near Northern Summer Solstice Observed by ACS/TIRVIM onboard TGO

Siteng Fan[1], Sandrine Guerlet[1,2], François Forget[1], Antoine Bierjon[1], Ehouarn Millour[1], Nikolay Ignatiev[3], Alexey Shakun[3], Alexey Grigoriev[4], Alexander Trokhimovskiy[3], Franck Montmessin[5], Oleg Korablev[3]

[1]LMD/IPSL, Sorbonne Université, PSL Research Université, École Normale Supérieure, École Polytechnique, CNRS, Paris, France
[2]LESIA, Observatoire de Paris, CNRS, Sorbonne Université, Université Paris-Diderot, Meudon, France
[3]Space Research Institute (IKI), Moscow, Russia
[4]Australian National University, Canberra, Australia
[5]LATMOS/IPSL, Guyancourt, France

**Contents of this file**

> Text S1 to S2
> Figures S1 to S3

**Introduction**

This supporting information includes the following content.

Text S1: a description of the LMD Mars GCM, which is used for simulating the diurnal temperature variations in the Martian atmosphere in this work.

Text S2 and Figure S1: an example of wave mode decomposition in the equatorial region at 100Pa.

Figure S2: wave mode decomposition results for the amplitudes of (0, 0) and (1, 0) modes, and the phases of (2, 2) and (1, -1) modes.

Figure S3: results of the (3, 3) mode amplitude in the test where the truncation of temporal frequency σ is expanded to ±3.

**Text S1.** LMD Mars GCM

The LMD Mars GCM (version 5.3) is used in this work for simulating the diurnal temperature variations in the Martian atmosphere (Fan 2022), which is one of the most popular Mars GCMs in this field for decades. A general description of the model architecture is given in Forget et al. (1999). It consists of a dynamical part and a physical part. The dynamical part solves the governing equations for atmospheric circulation on a three-dimensional grid, while the physical part contains parametrizations of physical processes in the Martian atmosphere which compute forcings in one-dimensional independent columns. Given the vertical range of TIRVIM observations from surface to ~1Pa, the thermosphere is not included in the GCM simulations used in our study. The vertical grid of the model contains 32 layers ranging from surface to ~$2\times10^{-3}$Pa, which is a hybrid coordinate linearly combined by topography and the sigma grid. The numbers of grid points along the longitude and the latitude dimensions are 64 and 48, respectively, which corresponds to 5.625° and 3.75°. The dynamical process has 960 steps on each Martian day (1.5 Martian minute/step), and the physical forcings are updated every 10 dynamical steps. Physical processes in the model include (1) radiative transfer with contributions from $CO_2$, clouds and dust, (2) cycles of $CO_2$, dust and water, and (3) transport of dust and gas, etc. The dust column-integrated opacity is normalized every Martian day at 14h to the MY 35 scenario built following Montabone et al. (2020). The output during $L_S$=60°-120° of MY 35 is available with this paper for further investigations.

**Text S2.** Example of Wave Mode Decomposition

Results of wave mode decomposition in the equatorial region between ±5° and 100Pa is shown in Figure S1 as an example. Observations in this latitude bin cover almost all local times and longitudes (Figure S1a), and consequently the fit successfully captures all variations (Figure S1b). The residuals are small (Figure S1c), mostly within 1-σ uncertainties, except for some outliers near ~3σ at 16h and 21h when the temperature is sampled at very large or small $L_S$ (Figure 1a).

In the mode decomposition, three wave modes show the largest amplitudes at almost all latitudes and pressure levels. The (1, 1) mode has the largest amplitude in most cases. At this pressure level (100Pa), it has an amplitude of ~4.0K and a crest near 9h (Figure S1d), which accounts for most of the temperature anomaly shown in Figure 3a. The (2, 2) mode shows an amplitude of ~2.6K and two crests near 4h and 16h (Figure S1e). However, same as that mentioned in Section 3.2, the derived semi-diurnal tide is possibly influenced by the TIRVIM spectra sampling scheme. Local times of the two wave crests coincide with those of the observations at large $L_S$ (Figure 1a), which amplifies the "apparent" semi-diurnal tide. The (1, -1) mode has a distinguishable contribution with an amplitude of 0.9K (Figure S1f). Although its amplitude is relatively smaller, this mode is visually noticeable in the temperature variation (Figure S1b). Results of other modes are computed but not shown here, as their contributions are much smaller.

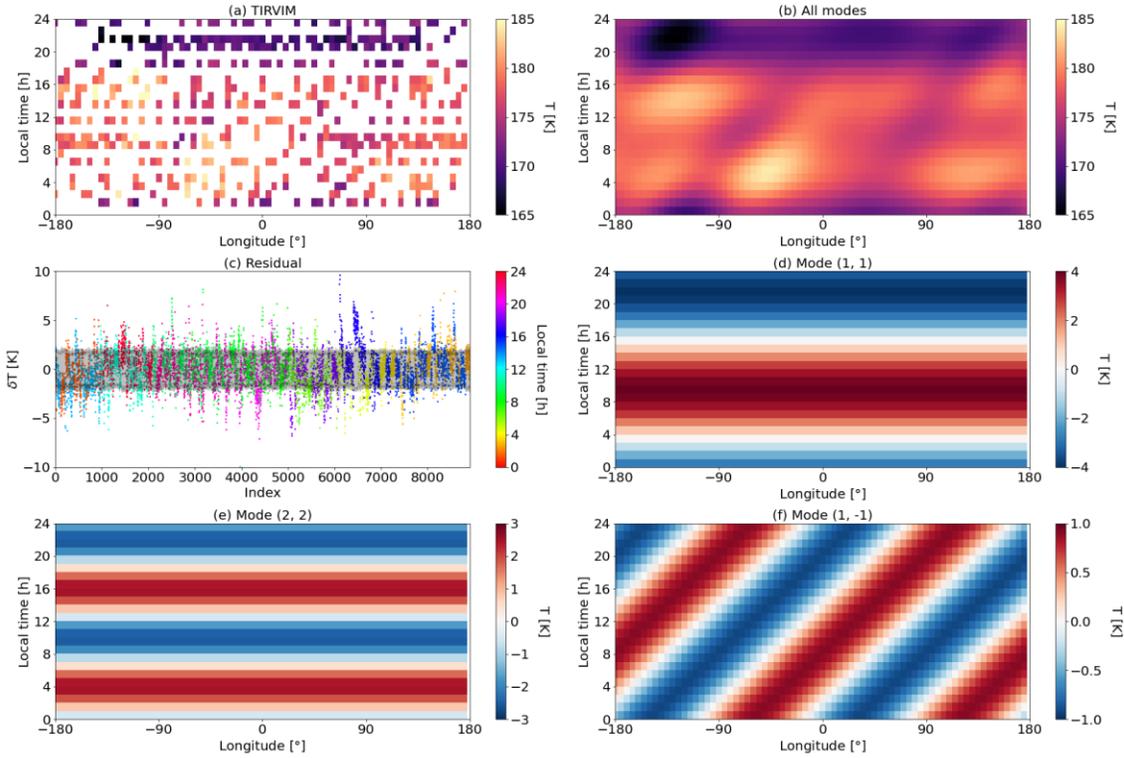

**Figure S1.** (a) Temperatures retrieved using TIRVIM observations at 100Pa, between ±5° in latitude during MY 35 L$_S$=75°-105°. (b) Wave mode decomposition results using the temperature data shown in (a). The decomposition is applied to the original individual temperature profiles and it uses universal time, while the results are presented in bins and local times for the purpose of illustration. (c) Residuals of the wave mode decomposition. Colors of the dots denote the local time, and the gray shaded area denotes the uncertainty derived from temperature retrieval. (d) The diurnal tide component, the (1, 1) mode, in the wave mode decomposition. (e) Same as (d), but for the semi-diurnal tide, the (2, 2) mode. (f) Same as (d), but for the diurnal Kelvin wave, the (1, -1) mode.

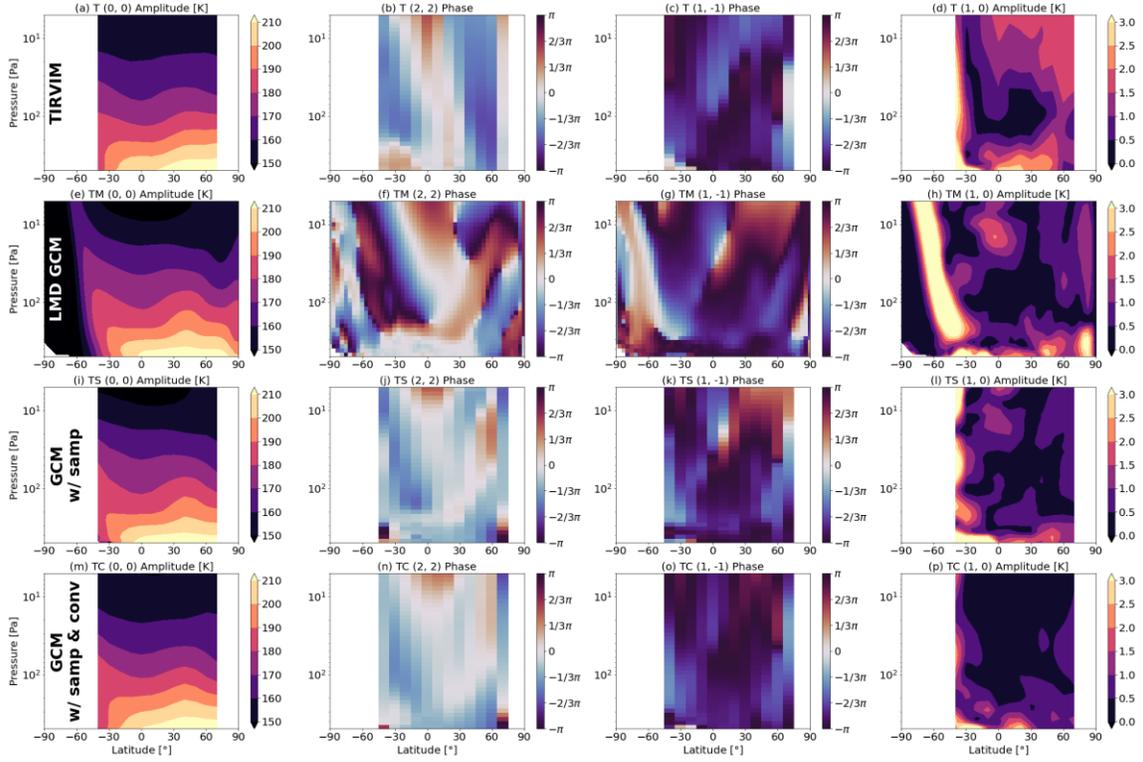

**Figure S2.** (a) Amplitude of the mean temperature, the (0, 0) mode, derived from the wave mode decomposition using TIRVIM observations. (b) Same as (a), but for the phase of the semi-diurnal tide. (c) Same as (b), but for the diurnal Kelvin wave. (d) Same as (a), but for the (1, 0) stationary wave. (e)-(h) Same as (a)-(d), respectively, but for the LMD Mars GCM outputs. (i)-(l) Same as (e)-(h), respectively, but for the model outputs sampled at the same locations and times as TIRVIM observations. (m)-(p) Same as (i)-(l), respectively, but it includes the effect of vertical convolution.

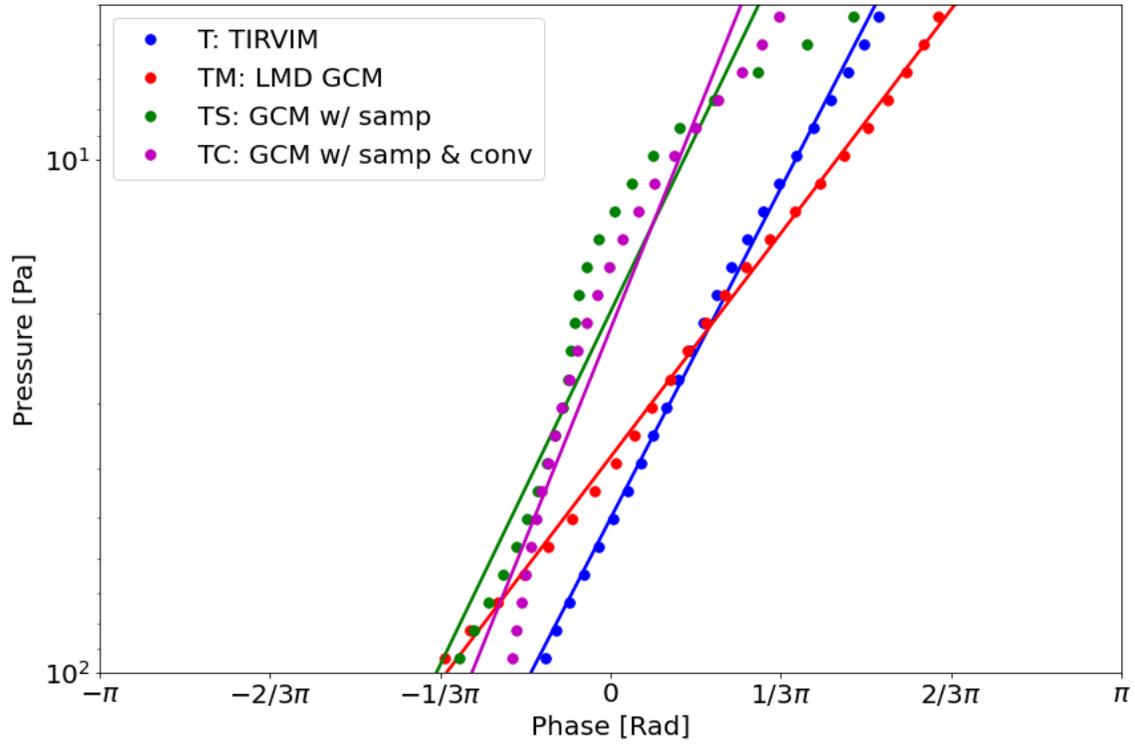

**Figure S3.** Phase of the semi-diurnal tide derived from the wave mode decomposition in the equatorial region using TIRVIM observations (blue dots), the LMD Mars GCM (red dots), the model outputs sampled at the same locations and times as TIRVIM observations (green dots), and the sampled model outputs including the effect of vertical convolution (magenta dots). They are the same values shown in color in the equatorial bins in Figure S2b, S2f, S2j, and S2n. Color lines are the linear fit results to the dots with the same colors, respectively.

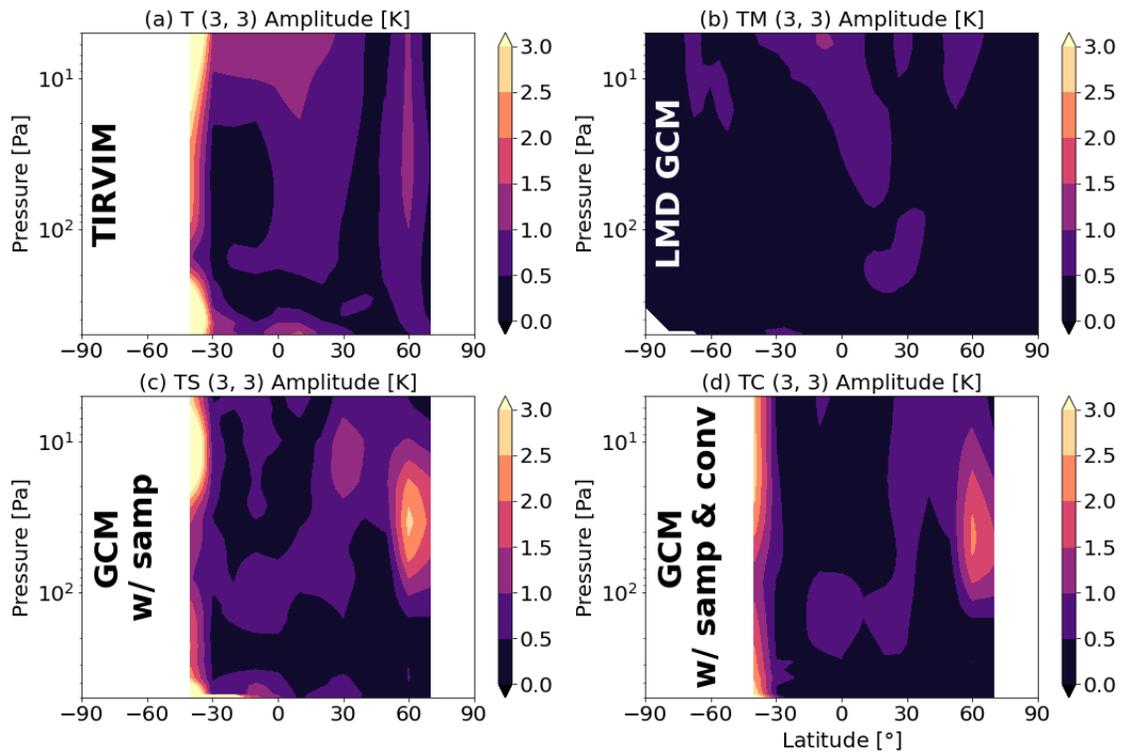

**Figure S4.** Amplitude of ter-diurnal tide, the (3, 3) mode, derived from test of wave mode decomposition, where the truncation of σ is expanded to ±3, using (a) TIRVIM observations, (b) the LMD Mars GCM outputs, (c) the model outputs sampled at the same locations and times as TIRVIM observations, and (d) the sampled model outputs including the effect of vertical convolution.